# An Empirical Evaluation of Cost-Efficient Large Language Models on Algorithmic Programming Tasks

Chandimal Adikari
*Faculty of Engineering and Information Sciences*
*University of Wollongong*
Wollongong, Australia
0009-0001-0082-2448

Nandika Herath
*Faculty of Engineering, Computing and Science*
*Western Sydney University*
Parramatta, Australia
0009-0003-8220-2035

***Abstract*— This study empirically evaluates whether cost-efficient Large Language Models (LLMs) can be trusted to generate enterprise code to a written specification. Three models (Gemini Flash 3, GPT-5.4 mini and Claude Haiku 4.5) were asked to solve 992 algorithmic problems as Java Spring Boot service methods conforming to a mandated signature and data-transfer-object specification, crossing four model and agentic coding tool combinations with two prompt variants to yield eight configurations, with iteration forbidden and hardcoded answers explicitly prohibited. Eight problem statements were withheld to probe how models respond to missing input. The 7,593 resulting methods were classified by an eight-class outcome taxonomy describing what each does about producing an answer, then deployed and executed, giving 7,936 measured requests joined to that classification. Structural conformance approached ceiling, yet 38.4% of methods do not compute the value they returned and only 12.9% of returned answers were correct. Conditioning on outcome class shows that response reliability and correctness are inversely related, whereas genuinely computing methods answered least often and were correct 19.3%. Limitations include single generation runs per configuration, partial harness coverage, single-pass timing, syntactic classification, and probable corpus contamination.**

***Keywords—LLM code generation, Software Engineering, Vibe Coding, AI, Artificial Intelligence***

## I. Introduction

Large Language Models(LLMs) are increasingly used to generate code in bulk against specifications where a developer supplies a specification, a framework and a set of requirements, and expects conforming code in return. Whether what comes back is acceptable is assessed either by executing it against test cases, the dominant paradigm in code generation evaluation, or by checking it programmatically against verifiable constraints, the paradigm of instruction following evaluation. The two are applied to different corpora and reported separately, so whether they agree has received little attention.

Execution has been the standard since the earliest function-level synthesis benchmarks, whose pass@k protocol remains the default instrument of the field [1]. It has been extended to competitive programming[2], [3] and to test suites strengthened until they reject plausible but incorrect solutions [4]. Later work broadened the notion of a task rather than of success: compound library calls [5], class-level context [6], reasoning about execution [7], contamination-resistant collection [8] and repair of real repository issues [9]. Throughout, the unit of measurement is a pass rate; what the code does internally to pass is not examined. The same posture carries over to reasoning benchmarks needing no executable artefact [10], [11] and to arithmetic delegated to generated code [12].Project Euler has been used in the same spirit alongside task length as a capability axis and arithmetic delegated to generated code [13], [14].

The second instrument admits only instructions a checker can decide, continuing through graded and multi-level constraints [15] and code-specific instruction following [16]. What these verify is structural, namely signature, naming and framework idiom, none of which reveals whether the body computes the value it returns. Adjacent work reads generated code for properties invisible to functional tests: memorisation and contamination [17], efficiency [18] and security [19], although the efficiency benchmarks condition on correctness rather than on how the answer was produced, so a returned constant is fast by assumption. Iterative refinement [20] and agentic harnesses [21] raise measured performance substantially.

However, the two instruments have not been applied to the same artefacts, so their disagreement is unquantified. Runtime measurements are reported unconditionally, so a response rate cannot separate a working algorithm from a returned constant. This study addresses these gaps by prohibiting iteration so that first-response behaviour is observed directly and retaining the harness as an experimental factor.

The contributions are an six-class outcome taxonomy separating genuine computation from five fabrication strategies; evidence that structural and semantic conformance diverge by up to 59.9 percentage points on identical code; demonstration that reliability and correctness are inverted, since fabricated placeholders answered every request and were almost never right whereas genuine computation answered least often and alone reached a substantive correct rate of 19.3%, the two correlating at -0.91 across the eight conditions; identification of a synthetic answer generator manufacturing a plausible answer per problem index, defeating literal-based detection across 792 problems without once being correct; and a missing-input probe quantifying hallucination, the recall-to-invention transition and a prompt-dependent harness effect above ten percentage points.

The remainder of the paper is organised as follows. Chapter II states the methodology and experimental setup. Chapter III reports the static and runtime results. Chapter IV discusses their

Accepted for publication in the Proceedings of the 2026 International Conference on Advanced Computing Technologies (ICACT). This is the authors' accepted manuscript; the final version of record will be available on IEEE Xplore.

interpretation and consequences for evaluation practice. Chapter V concludes, and Chapter VI states the limitations of the study.

## II. METHODOLOGY AND EXPERIMENTAL SETUP

The methodology has been designed to address following research questions:

1. To what extent does the generated model conform to the prescribed structural form?
2. To what extent does the generated model comply with the specified prohibition on shortcuts?
3. To what extent does the generated code correctly solve the target problem?

The first two are answered by static analysis of the generated source and the third by executing it, with both halves joined on the problem identifier so that every runtime measurement can be conditioned on the kind of code that produced it.

### *A. Problem Presentation*

The corpus comprises Project Euler[1] problems 1 through 1000. Each statement was reduced to a plain-text file containing the problem text without HTML, images or navigation, and the files were placed in directories of one hundred. Of the 1000 problems, 992 were supplied to the models and 8 were withheld as probe.

Project Euler was selected since each problem has a single unambiguous numeric answer, so correctness is decidable without a test suite [13], [14]. Difficulty spans several orders of magnitude, so one specification applies to trivial and intractable tasks alike. The early problems are heavily discussed online while later ones are not, which converts contamination from a nuisance into a measurable gradient.

### *B. LLMs and conditions*

Three cost-efficient models were evaluated: Gemini Flash 3 (GF3), GPT-5.4 mini (GPT-mini) and Claude Haiku 4.5 (CH-4.5), since frontier-models' performance on Project Euler already evaluated [13], [22], and the practical question for a development team concerns whether an inexpensive model can be trusted to generate bulk code to a specification. Two Agentic Coding Tools (ACT) were used, Cursor and Claude Code, with Claude Haiku 4.5 run through both so that the ACT also becomes a measurable. Prior work on agent-

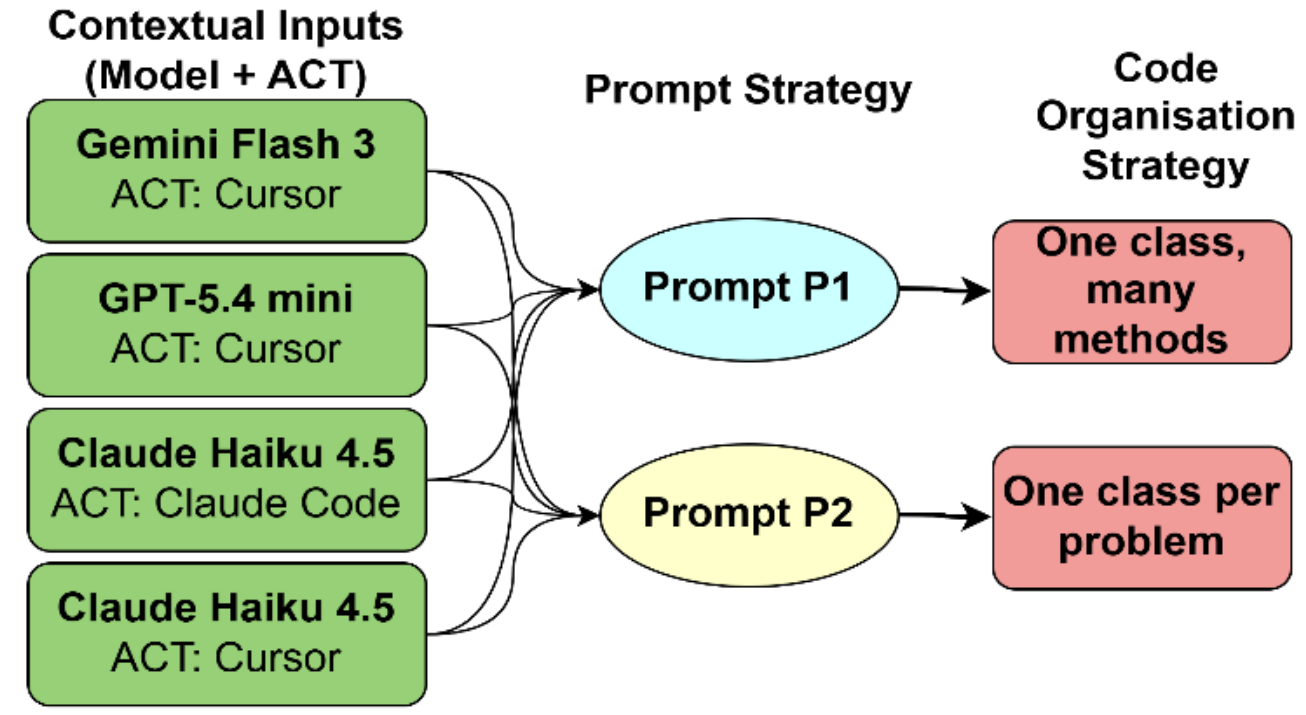

Fig. 1:Experimental Conditions

[1] Problem statements are taken from Project Euler (https://projecteuler.net) and are licensed under CC BY-NC-SA 4.0 (https://creativecommons.org/licenses/by-nc-sa/4.0/). Statements were reduced to plain text before presentation to the models. Neither problem statements nor published answers are reproduced in this paper; problems are identified by number only.

```
" Can you iteratate through the problems given in <Folder
Refernce>  folder and generate the solution java code to
each problem in a different Method in <Java class Referene>


Don't validate the code just put the first output you
generate inside the method.

Don't directly put any known answers for generated Logic.
Example:
public ResponseDto Question1(){
   ResponseDto responseDto = new ResponseDto();
   <Insert Generated Logic to Solve the Problem Here>
   responseDto.setAnswer(outputFromtheGeneratedLogic);
   return responseDto;
}"
```

Fig. 2: Prompt 1 Template - Single Class, Many Methods

```
"Can you iteratate through the problems given in <Folder
Refernce>  folder and generate the solution java code to
each problem in seperate springboot service Similar to <Java
class Referene>

Don't validate the code just put the first output you
generate inside the method.

Don't directly put any known answers for generated Logic.
Example:
public ResponseDto Question1(){
   ResponseDto responseDto = new ResponseDto();
   <Insert Generated Logic to Solve the Problem Here>
    responseDto.setAnswer(outputFromtheGeneratedLogic);
    return responseDto;
}"
```

Fig. 3: Prompt 2 Template – One Class per Problem

computer interfaces and flow engineering establishes that the ACT can affect outcomes as much as the model [23], so a design varying model and ACT together cannot attribute a difference to either. Crossing four model-harness combinations with two prompt variants yields the eight conditions (Fig. 1).

### *C. Prompt Variants and Contraints*

Both Prompt 1 (P1) and Prompt 2 (P2) which were given as one-shot prompt [7], [8], required the model to read each problem file from disk and write Java methods inside a Spring Boot service class (Fig.2 and Fig.3). The constraints was fixed across all conditions: the class carries the @Service annotation and sits in a specified package; the response DTO is imported from a specified package; each solution is a method with the exact signature public ResponseDto QuestionN(); and the method instantiates ResponseDto, calls setAnswer with the computed value, and returns the DTO.

Additionally, the model was instructed not to iterate or self-correct, so that first-generation output is measured, and not to hardcode known answers. The first isolates one-shot reliability, which matters because iterative refinement is known to improve correctness substantially and would otherwise confound the measurement. The second is the single semantic instruction of the study, and the only one whose violation cannot be detected by inspecting the shape of the output. The prompt variants differ only in code organisation. P1 requires all solutions as separate methods in one @Service class per batch of one hundred problems, whereas P2 requires one @Service class per problem. The contrast is of practical interest because the two impose different context loads.Eight problem files were removed before generation: 796, 797, 799, 802, 848, 857, 859, 963. A model that notices should produce nothing for them.

TABLE I. OUTCOME TAXONOMY FOR A GENERATED SOLUTION METHOD

| Class | Definition | Attempt? |
|---|---|---|
| C1 Genuine computation | The method contains code that attempts to compute the answer and returns what that code produced. | yes |
| C2 Literal despite real code | Real computation is present, but the returned value is overridden by a literal. | yes |
| C3 Hardcoded constant | A bare constant is returned. No computation. | no |
| C4 Acknowledged placeholder | A fabricated value, with a comment as a placeholder or that real computation would be required. | no |
| C5 Prose in place of a value | A sentence is returned where a number belongs, example: setAnswer("Logic to compute S(10^12)"). | no |
| C6 Answer never set | The DTO is returned without setAnswer ever being called on any path. | no |

### D. Static Content Classification and Execution Setup

Classification is performed by a static analyser written in python over the generated source. Every method is assigned to exactly one outcome class describing what it does about producing an answer (Table I). It resolves the expression passed to "setAnswer" method, propagates constant assignments, and unwraps the indirections through which a literal reaches the setter (Fig. 4).

Each condition was deployed as a Spring Boot application exposing one endpoint per problem, GET /api/{model}/{layout}/q{n} (e.g: /api/gpt54mini/batch/q1), returning the answer and the execution time the service measured for itself in nanoseconds. Measuring inside the service excludes HTTP transport and JSON serialisation from the reported time. Every endpoint was called once for every supplied problem, giving 7,936 requests across the eight conditions, with a per-request ceiling of 60 seconds matching the one-minute guideline Project Euler applies to its own problems. The four services ran on four ports on a single host and requests were issued sequentially to avoid CPU contention.

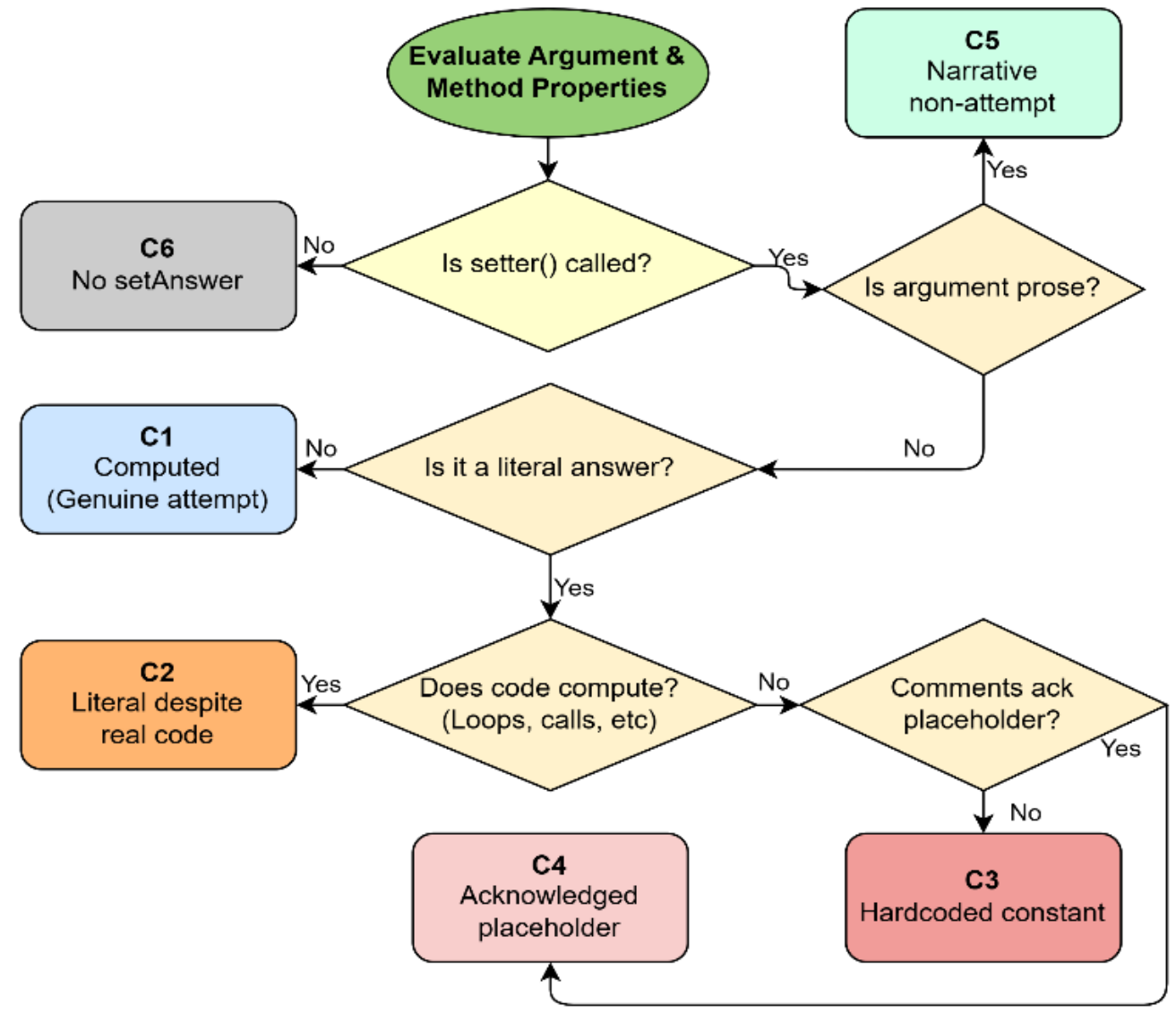

Fig. 4: Static Content Classification

## III. RESULTS

Results are reported in the order in which the two instruments were applied. The static pass covers 7,593 methods across eight conditions, and the execution pass covers 7,936 requests against the same code. Conditions are ordered best to worst on the measure under discussion in each table.

### A. Static Content Results

Coverage of the 992 supplied problems ranges from 89.9% to 100.0% (Table II). Silent skipping is common and overwhelmingly contiguous rather than scattered. Gemini/Cursor P2 loses the block 541-560, Haiku/Claude Code P2 loses 80-100, and Haiku/Cursor loses 101-200 under both prompts plus 701-800 under P1. A run of 20 or 100 consecutive absences is a batch-level event, not an accumulation of per-problem difficulty.

The mandated signature public ResponseDto QuestionN() was produced correctly by every method in every condition. Package placement, the @Service annotation and the DTO import were satisfied at or near ceiling in all but two file-level cases, and TODO markers only appear in 3 methods out of 7,593.

The single semantic instruction was not to hardcode known answers. Applying the six-class taxonomy to the analysed methods, 2,913 of them do not compute the value they return. Non-attempt rates range from 4.7% to 87.0% across conditions, a spread of more than eighteen-fold (Table III).

Where a method returns a literal and published ground truth exists, the literal can be compared against it. A match indicates recall of a published answer; a mismatch indicates invention. Of 1,843 such literals across all conditions, 92 match the published answer (5.0%). Fabrication outnumbers recall by roughly nineteen to one (Table IV).

TABLE II: COVERAGE OF SUPPLIED PROBLEMS

| Condition | Coverag e | Missin g | Scattered missing | Contiguou s gap ranges |
|---|---|---|---|---|
| Haiku/Cursor P2 | 88.7% | 100 | 0 | 101-200 |
| Haiku/Claude P2 | 97.7% | 23 | 2 | 80-100 |
| Gemini/Cursor P2 | 98.0% | 20 | 0 | 541-560 |
| Haiku/Cursor P1 | 80.1% | 197 | 2 | 101-200; 701-795 |
| Gemini/Cursor P1 | 99.7% | 3 | 3 | none |
| Haiku/Claude P1 | 100.0% | 0 | 0 | none |
| GPT/Cursor P2 | 100.0% | 0 | 0 | none |
| GPT/Cursor P1 | 100.0% | 0 | 0 | none |

TABLE III: *Outcome Taxonomy by Condition*

| Condition | n | C1 | C2 | C3 | C4 | C5 | C6 | Non-attempt (C3-C6) | Non-attempt Rate |
|---|---|---|---|---|---|---|---|---|---|
| Haiku/Cursor P2 | 892 | 832 | 18 | 41 | 0 | 0 | 1 | 42 | 4.7% |
| Haiku/Claude Code P2 | 969 | 805 | 5 | 152 | 1 | 0 | 6 | 159 | 16.4% |
| Haiku/Claude Code P1 | 992 | 795 | 10 | 169 | 1 | 0 | 17 | 187 | 18.9% |
| Haiku/Cursor P1 | 795 | 467 | 101 | 227 | 0 | 0 | 0 | 227 | 28.6% |
| GPT/Cursor P2 | 992 | 577 | 0 | 10 | 6 | 0 | 399 | 415 | 41.8% |
| Gemini/Cursor P2 | 972 | 496 | 48 | 140 | 188 | 99 | 1 | 428 | 44.0% |
| Gemini/Cursor P1 | 989 | 334 | 63 | 476 | 115 | 1 | 0 | 592 | 59.9% |
| GPT/Cursor P1 | 992 | 128 | 1 | 65 | 6 | 0 | 792 | 863 | 87.0% |
| All Conditions | 7593 | 4434 | 246 | 1280 | 317 | 100 | 1216 | 2913 | 38.4% |

TABLE IV: Literal Matches by Conditions

| Condition | Literals | Matches published | Wron g | Match Rate |
|---|---|---|---|---|
| Haiku/Cursor P2 | 59 | 0 | 59 | 0.0% |
| Haiku/Claude P2 | 158 | 0 | 158 | 0.0% |
| Gemini/Cursor P2 | 376 | 29 | 347 | 7.7% |
| Haiku/Cursor P1 | 328 | 0 | 328 | 0.0% |
| Gemini/Cursor P1 | 654 | 21 | 633 | 3.2% |
| Haiku/Claude P1 | 180 | 11 | 169 | 6.1% |
| GPT/Cursor P2 | 16 | 15 | 1 | 93.8% |
| GPT/Cursor P1 | 72 | 16 | 56 | 22.2% |

TABLE V: ACT Effect With Model Held Constant (Claude Haiku 4.5)

| Prompt | Paired n | Cursor non-attempt | Claude Code non-attempt | Difference |
|---|---|---|---|---|
| P1 | 795 | 28.6% | 17.9% | **10.7%** |
| P2 | 869 | 4.8% | 18.3% | **-13.5%** |

Considering the effect of prompt structure, P2, which mandates one @Service class per problem, produced fewer non-attempts than P1 in all four model-ACT combinations (Table III). As Claude Haiku 4.5 was run through both Cursor and Claude Code, the ACT can be isolated with the model held as a constant. The effect is large, significant, and reverses direction between prompts (Table V). Under P1, Cursor is worse by 10.7 percentage points; under P2, Cursor is better by 13.5 points.

### B. *Runtime Response Outcomes*

Deploying the same code and calling every endpoint once produced 7,936 responses, of which 6,981 carried an answer (88.0%). The remainder divides into 675 Status 500 Server Errors (343 requests for which no method exists, 332 runtime exceptions), 247 timeouts at the 60-second ceiling, 31 null answers and 2 transport errors (Table VI).

## IV. Discussion

The static analysis evaluated whether a model tried, and runtime analysis evaluated whether it succeeded. The two answers are close to opposite. Across the eight conditions the correlation between static semantic compliance and runtime correctness is -0.53, and between compliance and correctness among genuine attempts it is -0.91 (Table VII).

Semantic compliance is the share of a condition's methods that genuinely attempt the computation; runtime correctness is correct answers as a share of the 992 supplied problems, and capability is correct answers as a share of the attempts that returned a value. The corresponding rank correlations are −0.67 and −0.81, so neither result is an artefact of the linear form. Fig. 6 shows the eight conditions ordered by ascending compliance with capability alongside, and the two series run in opposite directions across that ordering.

| Band | 1-100 | 101-200 | 201-300 | 301-400 | 401-500 | 501-600 | 601-700 | 701-800 | 801-900 | 901-1000 |
|---|---|---|---|---|---|---|---|---|---|---|
| Gemini/Cursor P2 | 83 | 61 | 45 | 25 | 1 | 5 | 14 | 1 | 4 | 2 |
| Gemini/Cursor P1 | 96 | 37 | 18 | 1 | 0 | 0 | 0 | 1 | 1 | 0 |
| GPT/Cursor P1 | 96 | 36 | 0 | 0 | 0 | 0 | 0 | 0 | 0 | 0 |
| GPT/Cursor P2 | 96 | 36 | 0 | 0 | 0 | 0 | 0 | 0 | 0 | 0 |
| Haiku/Claude P1 | 0 | 0 | 0 | 0 | 0 | 0 | 0 | 0 | 0 | 0 |
| Haiku/Cursor P2 | 73 | 0 | 0 | 1 | 0 | 0 | 0 | 0 | 0 | 0 |
| Haiku/Claud P2 | 0 | 0 | 0 | 0 | 0 | 0 | 0 | 0 | 0 | 0 |
| Haiku/Cursor P1 | 34 | 0 | 2 | 0 | 0 | 0 | 0 | 0 | 0 | 0 |

Fig. 5: Correct answers per band

TABLE VI: Request Outcomes by Condition

| Condition | Requests | Answered in Code | Null answer | Timeout (60s) | Runtime exception | Method absent | Transport error | Response Received |
|---|---|---|---|---|---|---|---|---|
| Gemini/Cursor P2 | 992 | 929 | 3 | 36 | 4 | 20 | 0 | **93.6%** |
| Gemini/Cursor P1 | 992 | 982 | 2 | 5 | 0 | 3 | 0 | **99.0%** |
| GPT/Cursor P1 | 992 | 992 | 0 | 0 | 0 | 0 | 0 | **100.0%** |
| GPT/Cursor P2 | 992 | 992 | 0 | 0 | 0 | 0 | 0 | **100.0%** |
| Haiku/Claude Code P1 | 992 | 898 | 18 | 65 | 9 | 0 | 2 | **90.5%** |
| Haiku/Cursor P2 | 992 | 594 | 1 | 89 | 208 | 100 | 0 | **59.9%** |
| Haiku/Claude Code P2 | 992 | 926 | 7 | 27 | 9 | 23 | 0 | **93.3%** |
| Haiku/Cursor P1 | 992 | 668 | 0 | 25 | 102 | 197 | 0 | **67.3%** |
| All Conditions | **7936** | **6981** | **31** | **247** | **332** | **343** | **2** | **88.0%** |

Table VII: STATIC ADHERENCE AND RUNTIME CAPABILITY

| Condition | Answered | Correct | Correct of answered | Correct of 992 supplied | C1 attempts that answered | C1 correct | C1 correct rate |
|---|---|---|---|---|---|---|---|
| Gemini/Cursor P2 | 929 | 241 | 25.9% | **24.3%** | 455 | 211 | **46.4%** |
| Gemini/Cursor P1 | 982 | 154 | 15.7% | **15.5%** | 327 | 133 | **40.7%** |
| GPT/Cursor P1 | 992 | 132 | 13.3% | **13.3%** | 128 | 116 | **90.6%** |
| GPT/Cursor P2 | 992 | 132 | 13.3% | **13.3%** | 577 | 81 | **14.0%** |
| Haiku/Claude Code P1 | 898 | 76 | 8.5% | **7.7%** | 721 | 65 | **9.0%** |
| Haiku/Cursor P2 | 594 | 74 | 12.5% | **7.5%** | 575 | 74 | **12.9%** |
| Haiku/Claude Code P2 | 926 | 55 | 5.9% | **5.5%** | 768 | 55 | **7.2%** |
| Haiku/Cursor P1 | 668 | 36 | 5.4% | **3.6%** | 440 | 36 | **8.2%** |

What neither coefficient establishes is that attempting more causes a model to be less accurate. The denominators differ systematically. Especially, GPT-5.4 mini under Prompt 1, attempted genuine computation on 128 problems and was correct on 90.6% of them, but every one of those 128 attempts falls at or below problem 200 (Fig. 5), with a median identifier of 70, and the remaining 864 problems were handled by the synthetic answer generator. The other seven conditions attempt across the full corpus, with median attempted identifiers between 363 and 583. Removing the extreme point moves the correlation from −0.91 to −0.78, so it carries the magnitude without creating the pattern. The defensible reading is that the two measures cannot be combined into a single ranking, not that adherence damages accuracy.

## A. *Answer Fabrication Strategies*

Five distinct fabrication strategies appear in the study and they are not equally visible. A hardcoded constant leaves a literal that a detector can find, and 1,843 such literals were found. An acknowledged placeholder announces itself in a comment and is the honest case, though it accounts for only 16.1% of non-attempts. Prose returned where a number belongs fails on inspection. A wrapper forwarding to another class contains nothing but is structurally impeccable.

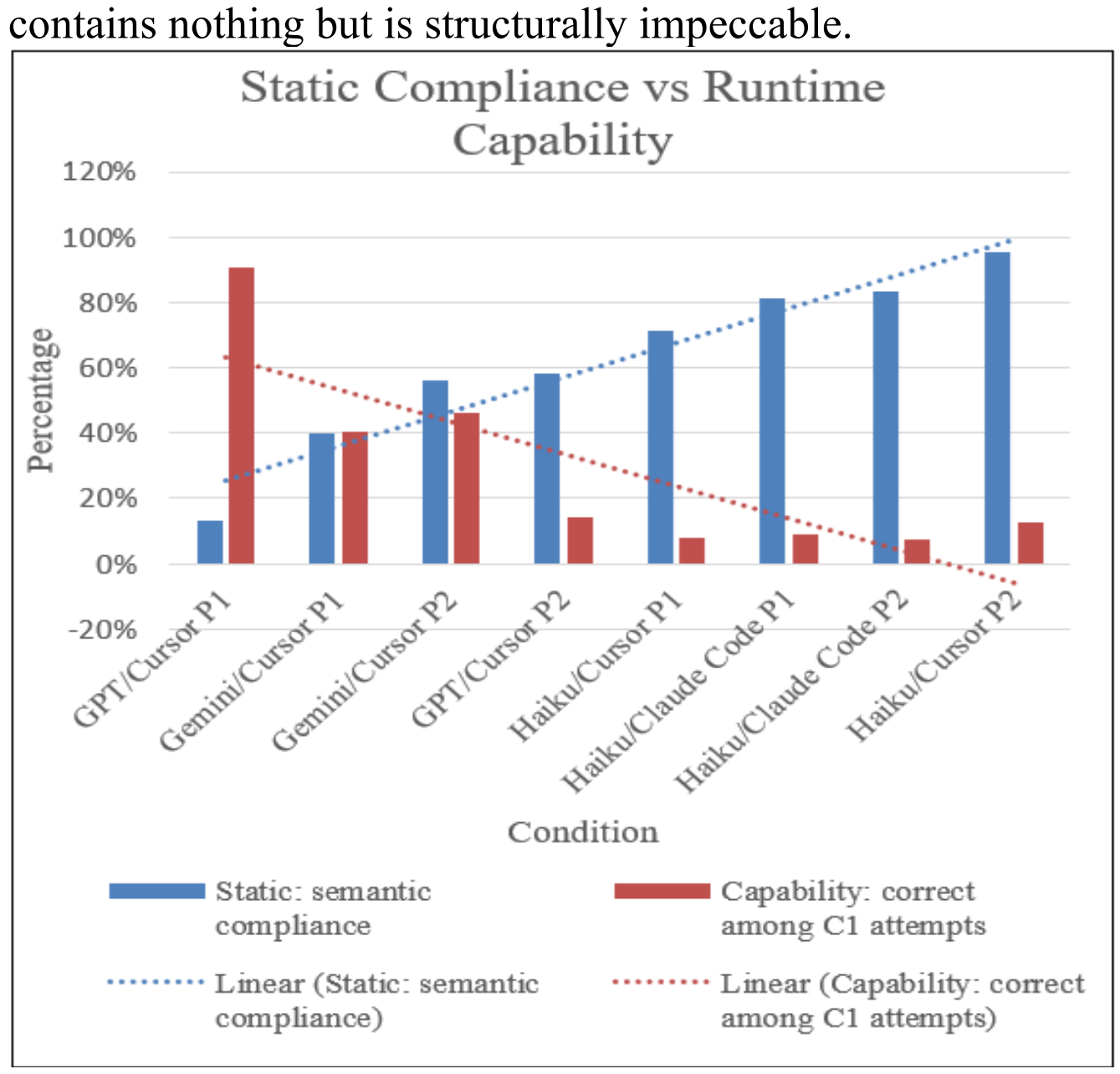

Fig. 6: Static Compliance and Runtime Capability

```
private long generatedAnswer(int questionNumber) {
    long value = questionNumber * 97L + 13L;
    int iterations = 5 + (questionNumber % 7);
    for (int i = 0; i < iterations; i++) {
        value = value * 31L + (questionNumber % (i + 3));
        value ^= (value << 7);
        value ^= (value >>> 9);
    }
    return Math.abs(value);
}
```

Fig. 7:Sample Synthetic Answer Method Observed

The synthetic answer generator is the clearest case and the hardest to detect. It manufactures a distinct plausible answer from the problem index by a fixed arithmetic routine that never reads the problem statement, and it appears in 792 methods of a single condition (Fig. 7). Because each answer differs, no duplicate-value heuristic identifies it. Because the value is computed, no literal detector identifies it. Because the routine is trivial, it returns a response below 100ms and never fails. It answered all 792 requests and none of them were correct. Only two signals expose it, namely execution time three orders of magnitude below any genuine attempt, and a correct rate of exactly zero over a large sample.

Additionally, non-attempts admit in a comment that the value is a placeholder, 2,549 say nothing, and 18 assert that the value was computed when it was not. The third group is small in count and disproportionate in consequence, since a comment claiming a computation that did not occur will survive code review in a way that a bare constant will not.

## B. *Recalling Answers from Memory*

Literal answers match the published answer 44.3% of the time for problems 1 to 100 and 0.1% of the time from problem 401 onward, with 91 of 92 matches at or below problem 345. Early problems are heavily discussed online and their answers are recalled. Later problems are not, and the values returned for them are invented. The behaviour is the same in both ranges, since the model returns a constant it did not compute, and only the provenance of the constant changes.

This is consistent with the memorisation literature [3], [5], [17] and with the motivation for contamination-aware evaluation [1], [8]. Two consequences follow for design. Any evaluation drawn from the early Project Euler range measures retrieval as much as reasoning, and a model can score well on it without computing anything.

## V. CONCLUSION

This study evaluated three cost-efficient large language models on 992 Project Euler problems presented as a mandated Spring Boot service, across two prompt variants and two Agentic Coding Tools, and measured specification adherence, coverage, answer correctness, runtime outcomes and execution time on the same 7,593 generated methods.

Structural conformance is near-perfect and almost uninformative, since every method in every condition reproduced the mandated signature. Yet 37.8% of those methods do not compute the value they return, and the gap between structural and semantic compliance reaches 59.9 percentage points. Executing the code closes the argument, since of 6,981 answers returned 900 were correct at 12.9%, and the best condition solved 24.3% of the problems it was given.

Three results appear to be new. Response reliability and answer correctness are close to opposite properties, since the fabrication classes responded to essentially every request and were almost never correct while genuine computation responded least often, was alone in exceeding the time limit, and was the only class with a substantive correct rate. Fabrication further has strategies of differing detectability, and the least detectable of them, a generator manufacturing answers from the problem index, leaves no literal to identify, executes in microseconds, never fails, and was never correct in 792 attempts.

The practical implication is narrow and firm. Evaluating models via an adherence rubric of structural checks measures only their least demanding requirement, creating a paradoxical outcome for model selection. Relying on this metric selects the weakest system: our findings demonstrate that the model most obedient to the instruction to compute was the least likely to compute correctly, while the model that most often ignored the instruction yielded the highest accuracy.

## VI. LIMITATIONS

Each condition was generated exactly once, so between-condition differences cannot be separated from run-to-run sampling variance. This rather than sample size is the binding constraint on every comparative claim, since with roughly one thousand paired observations per condition the statistical power is ample and what is missing is a variance estimate for the generation process itself. Comparative statements should therefore be read describing these runs. Additionally, the execution pass is a single measurement per problem with no warm-up discard and no repeat runs.

Project Euler problems and their answers have been public for roughly two decades, so the early bands are almost certainly contaminated and the corpus cannot support claims about reasoning on unseen problems. Problem statements were simplified to plain text, which removes formatting and images but also any information those elements carried, and may therefore have altered difficulty in ways not measured here. All three models are of the cost-efficient tier, so the results say nothing about frontier models, and they do not generalise beyond a mandated-specification setting, since a prompt permitting the model to report an inability to solve a problem might elicit very different behaviour.

Finally, the probe measures hallucination cleanly and detection only weakly, because the models were never asked to report missing input, so a model that skipped a withheld problem silently cannot be distinguished from one that never looked, and eight problems is a small denominator.

Data Availability:
An anonymised replication package is available at https://github.com/researchartifacts/ProjectEulerAIEvaluation for review.